\documentclass[fp,twocolumn]{jpsj3}
\usepackage{bm}
\usepackage{braket}
\usepackage{mathtools}

\usepackage{graphicx,amsmath,amssymb,color,nicefrac,multirow, makecell, ragged2e,float,multirow,arydshln,physics,ulem}

\newcommand{\non}{\nonumber\\}

\newcommand{\tx}[1]{{\text{#1}}}

\makeatletter
\providecommand*{\ext@figure}{lof}
\providecommand*{\ext@table}{lot}

\makeatother

\usepackage[numbers,sort&compress]{natbib}
\usepackage{url}
\usepackage{hyperref} 

\hypersetup{
    colorlinks=true, 
    citecolor=blue,  
    linkcolor=blue,  
    urlcolor=blue    
}

\title{\textbf{Scalar-spin-chirality-driven fractional Chern insulator on a kagome lattice}}

\author{Shinnosuke Tsutsumi\textsuperscript{1}, 
Koji Kudo\textsuperscript{1}, and 
Kentaro Nomura\textsuperscript{1,2}}
\inst{\textsuperscript{1}\textit{Department of Physics, Kyushu University, Fukuoka 819-0395, Japan}\\
\textsuperscript{2}\textit{Quantum and Spacetime Research Institute, Kyushu University, Fukuoka 819-0395, Japan}
} 

\abst{
  Fractional Chern insulators (FCIs) are the lattice analogs of the fractional 
  quantum Hall states, emerging even without an external magnetic field.
  In this work, we demonstrate the emergence of the FCI states in a kagome 
  magnet with a noncoplanar magnetic order that induces a finite scalar spin 
  chirality. By incorporating in our model both electron-electron interactions 
  and the effect of band dispersion, we find that stronger interactions
  relative to the band dispersion stabilize the FCI state over a broader range
  of scalar spin chirality. We characterize the emergent FCI state by 
  calculating overlap with representative states, identifying the ground-state degeneracy and
  the finite energy gap in the thermodynamic limit, and tracking the spectral flow under multiple flux-quantum insertions. 
  Our results suggest that kagome magnets with scalar spin chirality can be promising platforms for realizing FCIs.
}

\makeatletter
\def\@evenhead{} 
\def\@oddhead{}  
\makeatother

\begin{document}

\maketitle
\thispagestyle{plain} 

  \section{Introduction}

  The fractional quantum Hall effect~\cite{Tsui1982,Laughlin1983} is 
  one of 
  the most striking phenomena to arise from electron-electron interactions. 
  It exhibits fractionalized excitations that carry fractional charge and obey 
  fractional statistics 
  ~\cite{Arovas1984,Haldane1983,Halperin1984}, 
  which are the defining features of topologically ordered phases~\cite{Wen1995}. 
  Although the fractional quantum Hall effect requires a strong magnetic field,
  the resulting topological order stems from the nontrivial topology of the
  single-particle bands rather than the magnetic field itself. Fractional Chern
  insulators 
  (FCIs)~\cite{Sheng2011, Neupert2011, Regnault2011,Tang2011,Wu2012,Grushin2012} 
  realize such topological order without a magnetic field by partially filling 
  a nearly flat topological band that mimics Landau levels.
  In recent years, FCIs have been experimentally realized in moir\'e systems with
  nearly flat bands,
  such as twisted bilayer 
  $\text{MoTe}_2$~\cite{Cai2023,Park2023,Xu2023,Zeng2023,Ji2024,Redekop2024}
  and rhombohedral multilayer graphene on 
  hBN~\cite{Lu2024, Aronson2025, Lu2025,Xie2025}.
  Motivated by these developments, realizing FCIs in a broader range of
  material platforms beyond moir\'e systems has become an important direction.

  The kagome lattice is a typical example of a lattice model that hosts a flat 
  band. Together with Dirac dispersion, van Hove singularities, and geometric
  frustration, the kagome lattice has provided a fertile ground for
  diverse electronic and magnetic phenomena
  ~\cite{Liu2013,Sohal2018,Michal2021,Okamoto2022,Claassen2022,Sethi2023,Mallik2023,Comas2025,Fonseca2026,Lu2026,
  Guo2009,Balents2010,Kudo2017,Kudo2019,Tazai2023,Huang2024,Tazai2024,Shimura2024,
  Asaba2024,Nakazawa2024,Nakazawa2025,Nakazawa20252,Tazai20252,Huang2025,Onari2026,
  Goto2026,Yamakawa2026}, 
  including 
  superconductivity~\cite{Ortiz2021,zhao2021,Tazai2022,yang2023,Nagashima2025,Wang2025,Tazai2025,Kudo2026},
  the anomalous Hall effect~\cite{Ohgushi2000,Nakatsuji2015,Liu2018,Ye2018,Kobayashi2019,Fujiwara2023}, and various 
  magnetic structures~\cite{Sachdev1992,Han2012,Watanabe2022}.
  Experimentally, the kagome lattice has been realized in a wide range of 
  systems,
  such as Mn$_3$Sn~\cite{Nakatsuji2015,Ito2017,Liu2017,Kuroda2017,Higo2018,Higo20182}, Fe$_3$Sn$_2$~\cite{Ye2018,Lin2018,Fang2022},
  Mn$_3$Ge~\cite{Nayak2016,Kiyohara2016,Soh2020}, Co$_3$Sn$_2$S$_2$~\cite{Liu2018,Wang2018,Liu2019,Ozawa2019,Ikeda2021,Lau2023,Nakazawa20242}, 
  and CsV$_3$Sb$_5$~\cite{zhao2021,Ortiz2019,Ortiz2020,Tan2021,Murata2026}.
  Some of these materials, which we refer to as kagome magnets, spontaneously 
  break time-reversal symmetry due to magnetic ordering. When the resulting
  spin configuration is noncoplanar, the associated scalar spin chirality
  generates a topological band with nonzero Chern number, giving rise to a 
  Chern insulating state~\cite{Ohgushi2000}. This naturally raises a question 
  of whether analogous physics, combined with electron-electron interactions,
  could give rise to an FCI in kagome magnets.

  In this work, we demonstrate, using exact diagonalization, that the FCI state
  emerges in a wide region of scalar spin chirality. We construct an
  effective Hamiltonian projected onto a nearly flat topological band,
  incorporating repulsive electron-electron interactions along 
  with the effect of the band dispersion, to examine the stability of the 
  emergent FCI state. 
  We also confirm the ground-state degeneracy, the finite 
  energy gap in the thermodynamic limit, and the spectral flow with 
  multiple-flux quanta periodicity, which are characteristic signatures of the 
  FCI state. These results suggest that kagome magnets with a finite scalar 
  spin chirality are promising platforms for realizing FCIs.


  \begin{figure}[t]
    \centering
    \includegraphics[width=\linewidth]{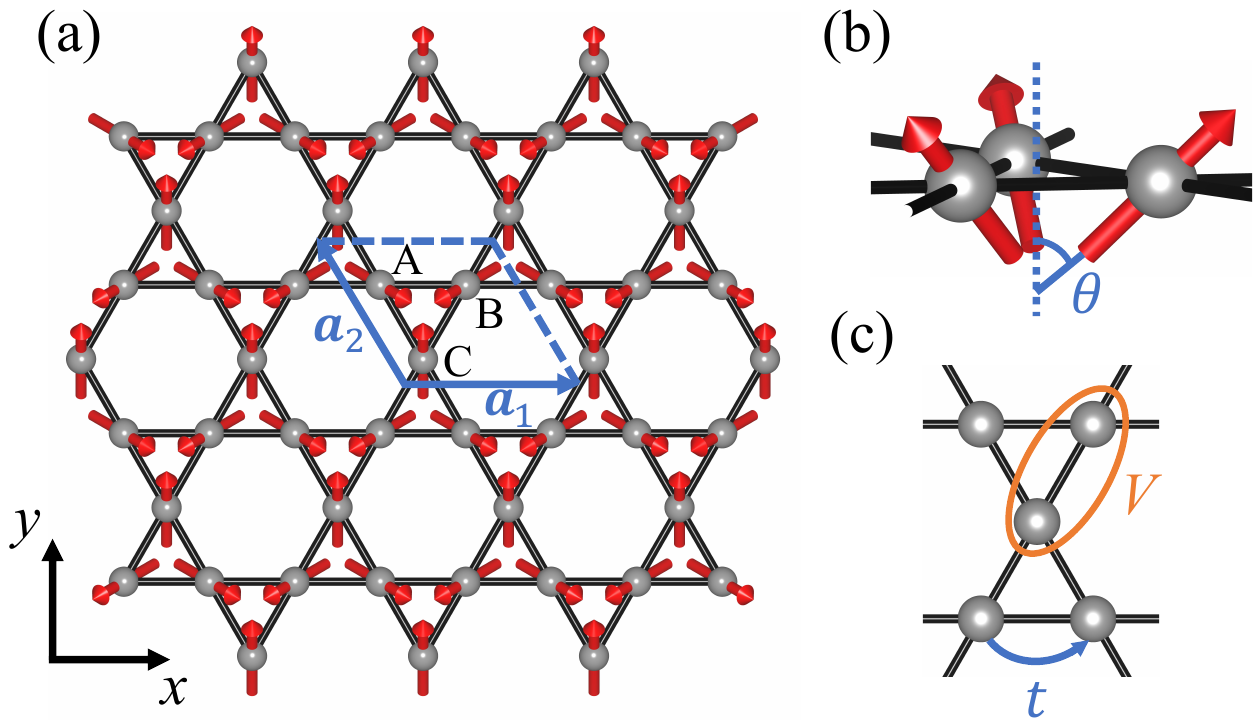}
    \caption{
    (a) Schematic illustration of a kagome lattice considered here. The red arrows represent
    localized spins. The blue rhombus represents a unit cell with sublattices A, B, 
   and C, and $\bm{a}_1$ and $\bm{a}_2$ are the lattice translation vectors.
   (b) Spin orientations specified by the common
    polar angle $\theta$. (c) System parameters appearing in $H_\tx{eff}$.
    }
    \label{model}
  \end{figure}

  \section{Model}
  \subsection{Spinful Hamiltonian}
  We consider a system of interacting spinful electrons on a kagome lattice,
  coupled to localized classical spins, see Fig.~\ref{model}(a). 
  Periodic boundary conditions are imposed.
  The Hamiltonian is given by 
  \begin{align}
    H_\tx{full}&=
    t\sum_{\langle ij\rangle}\sum_{\sigma}
    \tilde{c}_{i\sigma}^{\dagger}\tilde{c}_{j\sigma}
    -J_{\text{H}}S\sum_{i}\sum_{\sigma\sigma^{\prime}}
    \tilde{c}_{i\sigma}^{\dagger}
    (\bm{\sigma}\cdot\bm{n}_i)_{\sigma\sigma^{\prime}}
    \tilde{c}_{i\sigma^{\prime}}\non
    &\qquad\qquad
    +U\sum_i\tilde{n}_{i\uparrow}\tilde{n}_{i\downarrow} 
    +V\sum_{\langle ij\rangle}\sum_{\sigma\sigma^{\prime}}\tilde{n}_{i\sigma}
    \tilde{n}_{j\sigma^{\prime}},
  \end{align}
  where $\tilde{c}_{i\sigma}^{\dagger}$ is the creation operator of an electron
  with spin $\sigma$ on site $i$, 
  $\tilde{n}_{i\sigma}=\tilde{c}_{i\sigma}^{\dagger}\tilde{c}_{i\sigma}$, 
  $t$ is the nearest-neighbor hopping amplitude, $J_{\text{H}}$ is the 
  strength of the exchange coupling between electrons and a localized classical
  spin with the amplitude $S$ and the direction $\bm{n}_i$, and $\bm{\sigma}$ are the Pauli matrices.
  $U$ and $V$ are the strengths of the on-site and nearest-neighbor 
  interactions, respectively.
  We assume that the localized spins are aligned as the ``umbrella''
  structure as shown in Figs.~\ref{model}(a) and (b), 
  parameterized solely by a common polar angle $\theta$. 
  However, the following results apply to any spin configuration, 
  since emergent many-body states depend only on the solid angle subtended by the three spins 
  within a unit cell (i.e., the scalar spin chirality~\cite{Ohgushi2000}), as will be discussed below.

  
  \subsection{Strong exchange coupling limit}
  Hereafter, we consider the strong exchange coupling limit
  $J_{\text{H}}S/t \to \infty$. In this limit, electrons are fully 
  polarized along the direction of $\bm{n}_i$ at each site, and the system is
  effectively described by a spinless model. The creation operator of the 
  corresponding electron is defined by
  \begin{equation}
    c_i^{\dagger}
     = (\tilde{c}_{i\uparrow}^{\dagger},
    \tilde{c}_{i\downarrow}^{\dagger})\bm{\chi}_i,
  \end{equation}
  where $\bm{\chi}_i$ is a two-component spinor that satisfies
  $(\bm{\sigma}\cdot\bm{n}_i)\bm{\chi}_i=\bm{\chi}_i$. It is explicitly given
  by
  \begin{equation}
    \bm{\chi}_i = 
      \begin{pmatrix}
      \cos{(\theta_i/2)}\\
      e^{i\phi_i}\sin{(\theta_i/2)}
    \end{pmatrix},
  \end{equation}
  where $\theta_i$ and $\phi_i$ are the polar coordinates of $\bm{n}_i$.

  By projecting onto the spin-polarized subspace realized in the limit $J_{\text{H}}S/t \to \infty$, i.e., performing the 
  transformation $(\tilde{c}_{i\uparrow},\tilde{c}_{i\downarrow})
  \rightarrow c_i^\dagger\bm{\chi}_i^\dagger$,
  an effective spinless Hamiltonian is given by
  \begin{align}
    H_\tx{spinless}
    = t\sum_{\langle ij\rangle}c_{i}^{\dagger}\bm{\chi}_i^{\dagger}
    \bm{\chi}_jc_{j} + V\sum_{\langle ij\rangle}n_{i}n_{j},
  \end{align}
  where $n_i = c_i^{\dagger}c_i$. Here, a constant arising from the exchange 
  coupling is neglected and the on-site interaction vanishes due to the Pauli 
  principle. The factor 
  $\bm{\chi}_i^{\dagger}\bm{\chi}_j$ generates a complex hopping 
  amplitude, leading to effective magnetic fluxes through closed loops. The
  accumulated phase around a loop is determined by the scalar spin chirality
  of the three spins~\cite{Ohgushi2000}.
  In numerical calculations, we restrict the polar angle $\theta$ to $0 \leq \theta \leq \pi/2$, since the systems with 
  $\theta$ and $\pi - \theta$ yield the same magnitude of scalar spin chirality with opposite signs, 
  resulting in the same energy spectra.
  
  In the momentum space, the Hamiltonian becomes
  \begin{align}
   \label{Fourier}
    H_\tx{spinless} 
   &= \sum_{\bm{k}}\bm{c}_{\bm{k}}^{\dagger}h_0(\bm{k})\bm{c}_{\bm{k}}\non
    &+ \sum_{\bm{k}_1\bm{k}_2\bm{k}_{3}\bm{k}_{4}}\sum_{(\alpha_1,\alpha_2)}
    V_{\bm{k}_1\bm{k}_2\bm{k}_{3}\bm{k}_{4}}^{\alpha_1\alpha_2}
    c_{\bm{k}_1\alpha_1}^{\dagger}c_{\bm{k}_2\alpha_2}^{\dagger}
    c_{\bm{k}_{3}\alpha_2}c_{\bm{k}_{4}\alpha_1},
  \end{align}
  where $\bm{c}_{\bm{k}}^{\dagger}
  =(c_{\bm{k}\text{A}}^{\dagger},c_{\bm{k}\text{B}}^{\dagger},
  c_{\bm{k}\text{C}}^{\dagger})$ is the vector of the creation operators with 
  the subscripts A, B, and C labeling the sublattices, 
  and $\sum_{(\alpha_1,\alpha_2)}$ denotes the summation
  over the pairs $(\text{A},\text{B})$, $(\text{B},\text{C})$, and 
  $(\text{C},\text{A})$.
  The matrix $h_0(\bm{k})$ takes the form of
  \begin{equation}
    h_0(\bm{k}) = t
    \begin{pmatrix}
      0&\chi_{\text{AB}}(1 + e^{-i\bm{k}\cdot\bm{a}_1})
      &\chi_{\text{AC}}(1 + e^{-i\bm{k}\cdot\bm{a}_3})\\
      & 0 & \chi_{\text{BC}}(1 + e^{-i\bm{k}\cdot\bm{a}_2})\\
      \tx{h.c.} & & 0
    \end{pmatrix},
  \end{equation}
  where $\bm{a}_1$ and $\bm{a}_2$ are the lattice translation vectors as shown in Fig.~\ref{model}(a), 
  $\bm{a}_3=\bm{a}_1+\bm{a}_2$, and $\chi_{\alpha\beta} = \bm{\chi}_{\alpha}^{\dagger}\bm{\chi}_{\beta}$.
  The interaction matrix elements are
  \begin{align}
    V_{\bm{k}_1\bm{k}_2\bm{k}_3\bm{k}_4}^{\alpha_1\alpha_2} 
    &= \frac{V}{N_1N_2}
    \delta_{\bm{k}_1+\bm{k}_2-\bm{k}_3-\bm{k}_4,\bm{0}}^{\text{mod} \bm{N}}\non
    \times&
    \begin{cases}
      1 + e^{-i(\bm{k}_1-\bm{k}_4)\cdot\bm{a}_1} & (\alpha_1,\ \alpha_2) = (\text{A, B}) \\
      1 + e^{-i(\bm{k}_1-\bm{k}_4)\cdot\bm{a}_2} & (\alpha_1,\ \alpha_2) = (\text{B, C}) \\ 
      1 + e^{-i(\bm{k}_2-\bm{k}_3)\cdot\bm{a}_3} & (\alpha_1,\ \alpha_2) = (\text{C, A}) \\
    \end{cases}.
  \end{align}
  Here, $N_i$ is the number of unit cells along the $\bm{a}_i$ direction, and  
  $\delta_{\bm{k},\bm{k}'}^{\text{mod}\bm{N}}= \delta_{j_1,j_1'}^{\text{mod}N_1}\delta_{j_2,j_2'}^{\text{mod}N_2}$,
  where $\delta_{j,j'}^{\text{mod} N} = 1$ if $j \equiv j'\pmod{N}$, and $0$ 
  otherwise. 
    
  \subsection{Projection onto the lowest band}
  We now consider the regime where the band filling, $\nu=N_e/(N_1N_2)$ 
  ($N_e$: the electron number), is less
  than one, and the interaction strength is smaller than the band gap.
  Following the standard treatment in the FQH problem, we construct an 
  effective Hamiltonian projected onto the lowest band as
  \begin{align}
    H_{\text{eff}} &= 
    \sum_{\bm{k}}\varepsilon(\bm{k})d_{\bm{k}}^{\dagger}d_{\bm{\bm{k}}}
    + \sum_{\bm{k}_1\bm{k}_2\bm{k}_3\bm{k}_4}
    V_{\bm{k}_1\bm{k}_2\bm{k}_3\bm{k}_4}
    d_{\bm{k}_1}^{\dagger}d_{\bm{k}_2}^{\dagger}d_{\bm{k}_3}d_{\bm{k}_4},
    \label{eq:Heff}
  \end{align}
  where $d_{\bm{k}}^{\dagger}=\bm{c}_{\bm{k}}^{\dagger}\bm{\psi}_{\bm{k}}$ and
  $\bm{\psi}_{\bm{k}}=(
  \psi_{\bm{k}\tx{A}},\psi_{\bm{k}\tx{B}},\psi_{\bm{k}\tx{C}})^T$
  is the eigenvector of $h_0(\bm{k})$ with the lowest eigenvalue 
  $\varepsilon(\bm{k})$. Note that we retain the kinetic term in $H_\tx{eff}$
  since the bandwidth is finite, unlike the Landau level. The pseudopotentials 
  are given by 
  \begin{align}
    V_{\bm{k}_1\bm{k}_2\bm{k}_3\bm{k}_4} &= \sum_{(\alpha_1,\alpha_2)}
    V_{\bm{k}_1\bm{k}_2\bm{k}_3\bm{k}_4}^{\alpha_1\alpha_2}
    \psi_{\bm{k}_1\alpha_1}^{\ast}\psi_{\bm{k}_2\alpha_2}^{\ast}
    \psi_{\bm{k}_3\alpha_2}\psi_{\bm{k}_4\alpha_1}.
  \end{align}
  At this stage, the system is governed by three parameters: the polar angle
  $\theta$ of spins, the hopping amplitude $t$, and the interaction strength $V$, 
  as summarized in Figs.~\ref{model}(b) and (c).

  In the following, we perform exact diagonalization of $H_{\text{eff}}$ 
  to investigate the emergence of FCI states at $\nu=1/3$.
  We treat a finite system consisting of $N_1\times N_2$ unit cells.
  For a given $N_e$, we choose $(N_1, N_2)$ such that the aspect ratio $N_1/N_2$ is
  closest to unity, with $N_1\geq N_2$. 
  As the Hamiltonian conserves the total momentum $\bm{K}$,
  we diagonalize $H_{\text{eff}}$ using the Lanczos algorithm within each $\bm{K}$ sector. 
  Here, the momentum is discretized as $\bm{K} = (K_1, K_2)$, where $K_{i} = \bm{K}\cdot\bm{a}_i=2\pi j_{i}/N_{i}$ 
  with the momentum index $j_i = 0, 1, \cdots, N_{i}-1$.

  \begin{figure}[t]
    \centering
    \includegraphics[width=\columnwidth]{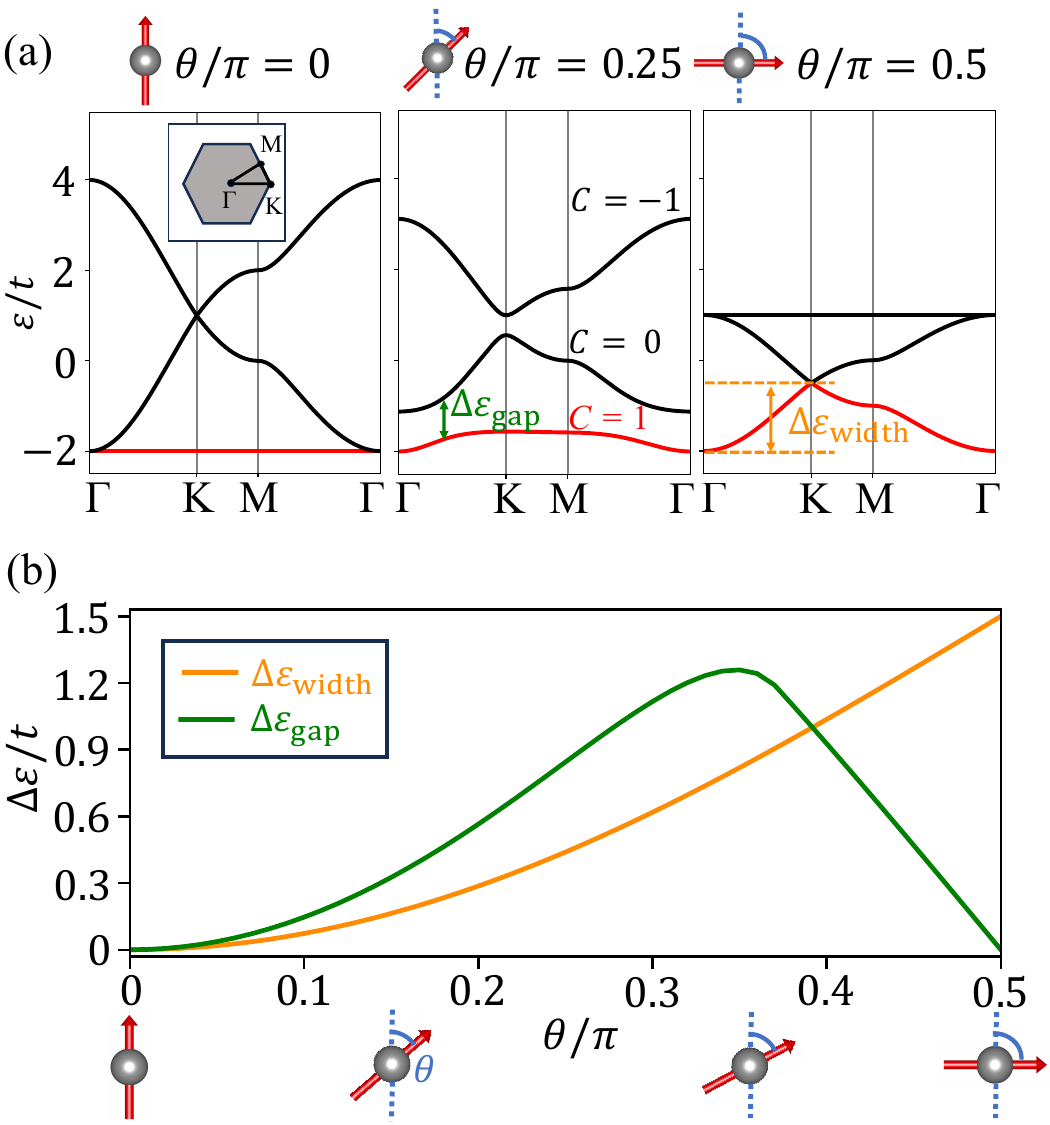}
    \caption{
      (a) Band structures of the non-interacting Hamiltonian $h_0$ for different
      polar angles $\theta$. For $0 < \theta/\pi < 0.5$, all bands are isolated, 
      and the band Chern numbers from bottom are $1$, $0$, and $-1$, respectively.
      (b) Bandwidth of the lowest band $\varepsilon_{\text{width}}$ and its direct
      energy gap $\varepsilon_{\text{gap}}$ as functions of $\theta$.  
   Both $\varepsilon_{\text{width}}$ and $\varepsilon_{\text{gap}}$ become 
   finite and are of comparable magnitude as $\theta$ increases.
    }
  \label{band}
  \end{figure}
  
  \section{Numerical results}
  \subsection{Band structure}
  Before discussing the many-body problem, we briefly review the Chern insulating state
  realized in the kagome magnet~\cite{Ohgushi2000}.
  Figure~\ref{band}(a) shows the energy bands of $h_0(\bm{k})$ for different
  polar angles $\theta$. For any $\theta$ except $\theta=0$ and $\pi/2$, all 
  three bands are isolated from each other, and their band Chern numbers from 
  the bottom are $1$, $0$, and $-1$, respectively.

  While the lowest band is completely flat at $\theta=0$, it acquires 
  dispersion as $\theta$ increases. To quantify this, we plot in 
  Fig.~\ref{band}(b) the bandwidth of the lowest band and its direct band gap.
  This indicates that the two energy scales are comparable, implying that
  the band dispersion can play a non-negligible role when the interaction 
  strength is smaller than the band gap. This motivates us to include the 
  kinetic term in $H_\tx{eff}$ in Eq.~\eqref{eq:Heff}.
  
  \begin{figure}[t]
   \centering
   \includegraphics[width=\columnwidth]{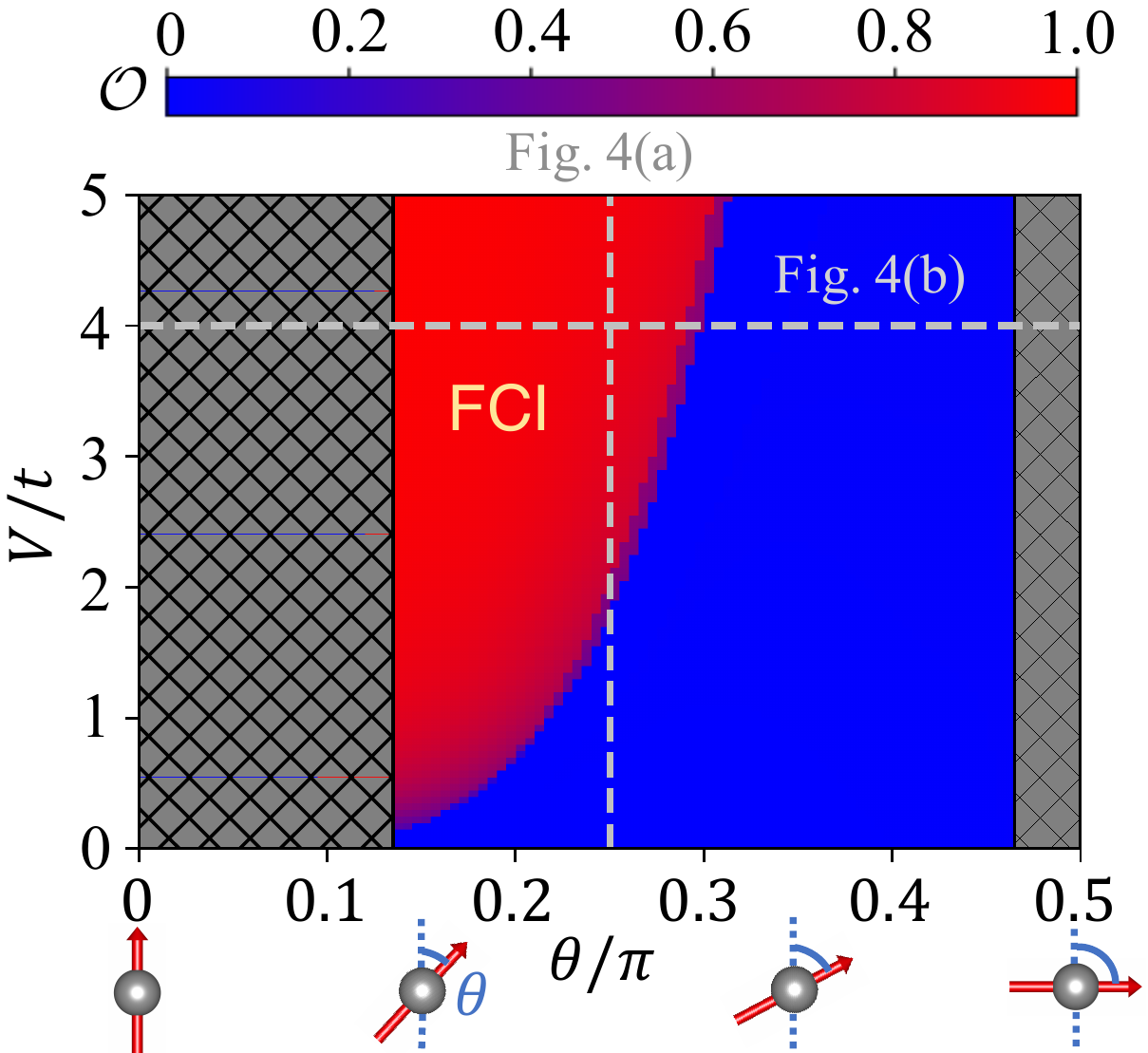}
   \caption{
   Average of the singular values of the overlap matrix $O$ as a function of the polar angle $\theta$ and 
   interaction strength $V$. The red region, where $\mathcal{O}$ is close to unity, indicates the emergence of the FCI state.
   The hatched gray region indicates that the three lowest energy states in the flat-band limit do not match the counting rule
   of the total momentum in Eq.~\eqref{fci_momentum}.
   The system size is $N_1 \times N_2 = 6 \times 4$ and the band filling 
   factor is $\nu= 1/3$.
   }
   \label{overlap}
  \end{figure}

  \begin{figure}[t]
    \centering
    \includegraphics[width=\columnwidth]{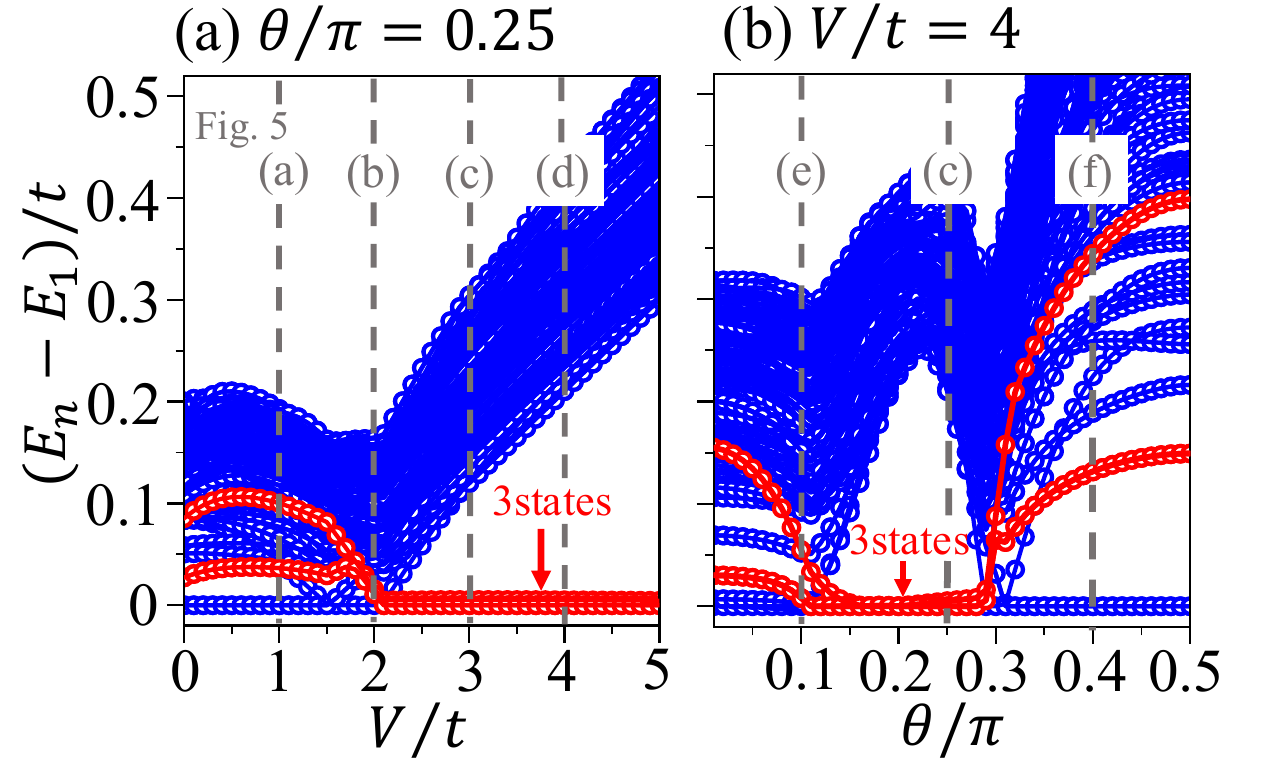}
    \caption{
   Low-lying many-body energy spectra $E_n-E_1$, where $E_n$ is the $n$-th 
   lowest energy, along the two cuts indicated in Fig.~\ref{overlap}.
   The lowest energy states in each momentum sector
    $(j_1,j_2) = (0,0), (2,0), (4,0)$ [Eq.~\eqref{fci_momentum}] are 
   highlighted in red. The three-fold quasidegenerate states appear in
   the regions with $\mathcal{O}\approx1$ in Fig.~\ref{overlap}.
   We plot six lowest energies in each momentum sector.
   }
    \label{vint_theta_spectrum}
  \end{figure}
  \subsection{Scalar-spin-chirality-driven fractional Chern insulator}
  We now move on to the interacting problem. Our main result is summarized in
  Fig.~\ref{overlap}, which shows the parameter region $(\theta,V/t)$ where
  the FCI state is expected to emerge. This figure is generated as follows:
  According to Refs.~\cite{Wu2012,Tang2011}, the
  $\nu=1/3$ FCI state emerges in the simplest model on a kagome lattice. 
  This model is identical to our system at $\theta/\pi \simeq 0.31$
  in the flat-band limit $V/t\rightarrow\infty$. 
  The total momentum $\bm{K} = 2\pi(j_1/N_1, j_2/N_2)$ for these FCI states are
  given by~\cite{Regnault2011}
  \begin{align}
   \begin{split}
    j_1&=\frac{N_1N_2(N_1 + 2s -3)}{6}\  (\tx{mod } N_1),\\
    j_2&=\frac{N_1N_2(N_2-1)}{6}\ (\tx{mod } N_2),
   \end{split}
   \label{fci_momentum}
  \end{align}
  where $s = 0, 1, 2$ labels the three-fold (quasi)degeneracy.
  Here, we assume that $N_1$ is a multiple of three. 
  [If this is not the case, the subscripts 1 and 2 in Eq.~\eqref{fci_momentum} are to be interchanged.]
  Using these flat-band limit states as representative FCI states, we 
  define the overlap matrix $O$ at given parameters $(\theta,V/t)$ as
  \begin{align}
   O_{ij}=\bra{\Psi_i(\theta,\infty)}\ket{\Psi_j(\theta,V/t)} 
   (i,j=1,2,3),
  \end{align}
  where $\ket{\Psi_i(\theta,V/t)}$ is the $i$-th lowest energy eigenstate for 
  $(\theta,V/t)$.
  The average of the singular values of $O$, denoted by $\mathcal{O}$, serves as
  an indicator of the overlap between two degenerate states~\cite{Geraedts2015}:
  $\mathcal{O}=1$ when they coincide and $\mathcal{O}=0$ when they are 
  orthogonal.
  
  Figure~\ref{overlap} plots $\mathcal{O}$ as a function of $\theta$ and $V/t$.  
  We observe a broad red region with $\mathcal{O}\approx1$, indicating that the FCI state emerges
  in the kagome magnets considered here.
  Notably, the FCI state emerges over a wider range of scalar spin
  chirality as the interaction strength $V$ is increased.
  The hatched gray region indicates that we do not have representative states since the three lowest states in the flat-band
  limit do not appear in the momentum sectors in Eq.~\eqref{fci_momentum}. (See Appendix for more details.)

  \begin{figure}[t]
   \centering
   \includegraphics[width = \columnwidth]{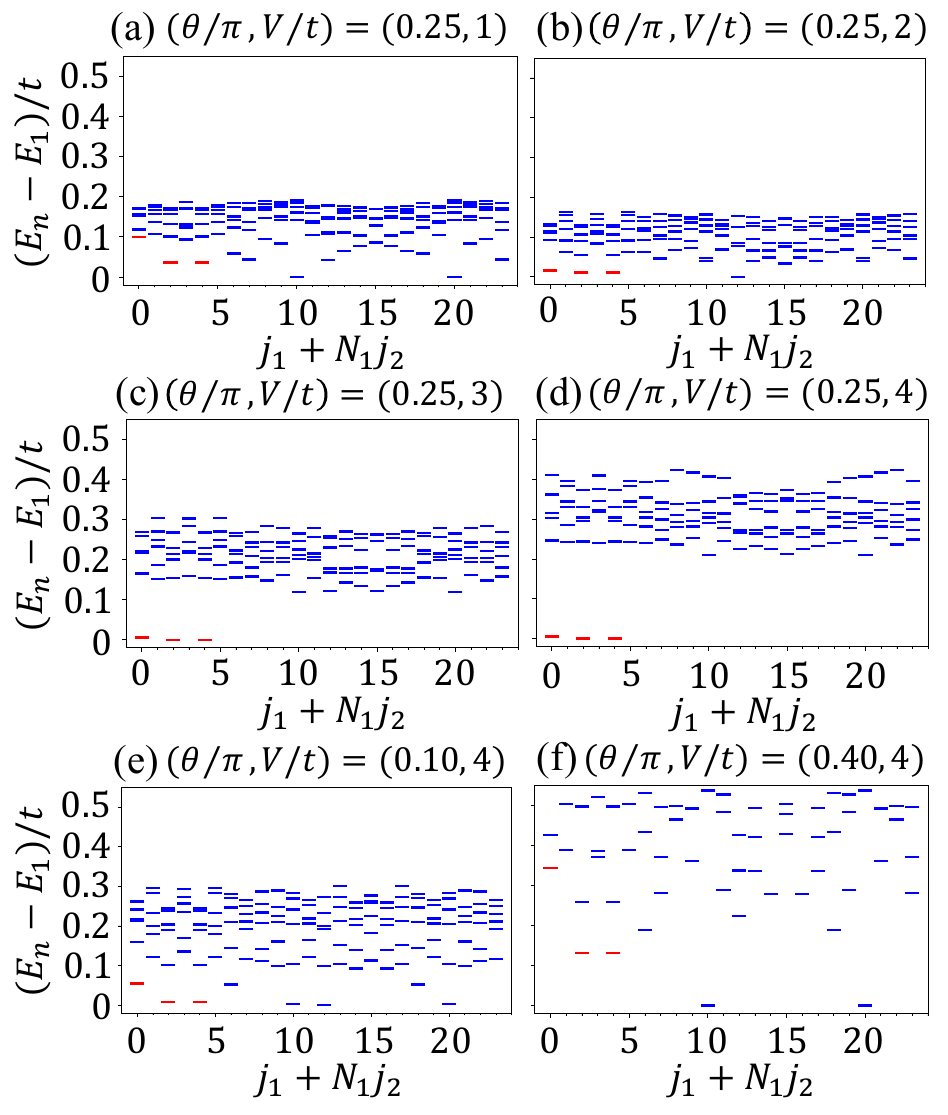}
   \caption{
   Low-lying many-body spectra as functions of the total momentum index 
   $j_1 + N_1j_2$ for the parameters shown in
   Fig.~\ref{vint_theta_spectrum}. 
   The red markers indicate the lowest energies in each momentum sector 
   $(j_1, j_2) = (0,0), (2,0), (4,0)$ [Eq.~\eqref{fci_momentum}].
   We plot the six lowest energies in each momentum sector.
   }
   \label{ene_spe}
  \end{figure}
  Now, we present further numerical results regarding the ground-state degeneracy and the energy gap. 
  Figure~\ref{vint_theta_spectrum} shows the many-body spectra $E_n-E_1$ along the two cuts indicated in 
  Fig.~\ref{overlap}, where $E_n$ is the $n$-th
  lowest energy. We highlight in red the lowest energy states in each momentum
  sector given by Eq.~\eqref{fci_momentum}. In Figs.~\ref{vint_theta_spectrum}(a) and
  (b), the three-fold quasidegenerate ground states carrying these momenta 
  clearly emerge in the ``FCI'' regions identified in Fig.~\ref{overlap}. 
  Here, we refer to almost degenerate states separated from 
  higher excited states by a finite gap as quasidegeneracy.
  
  Figure~\ref{ene_spe} shows many-body spectra as functions of the total 
  momentum index $j_1 + N_1j_2$ at various
  values of $\theta$ and $V/t$ as indicated in Fig.~\ref{vint_theta_spectrum}.
  Figures~\ref{ene_spe}(a)-(d) share the same $\theta/\pi=0.25$, where we 
  confirm the emergence and evolution of the three-fold quasidegenerate 
  ground state as $V/t$ increases. As shown in Fig.~\ref{ene_spe}(e) and (f),
  the three-fold quasidegenerate ground state moves up in energy when 
  $\theta$ is close to $0$ or $\pi/2$.
  
  \begin{figure}[t]
   \centering
   \includegraphics[width=1\columnwidth]{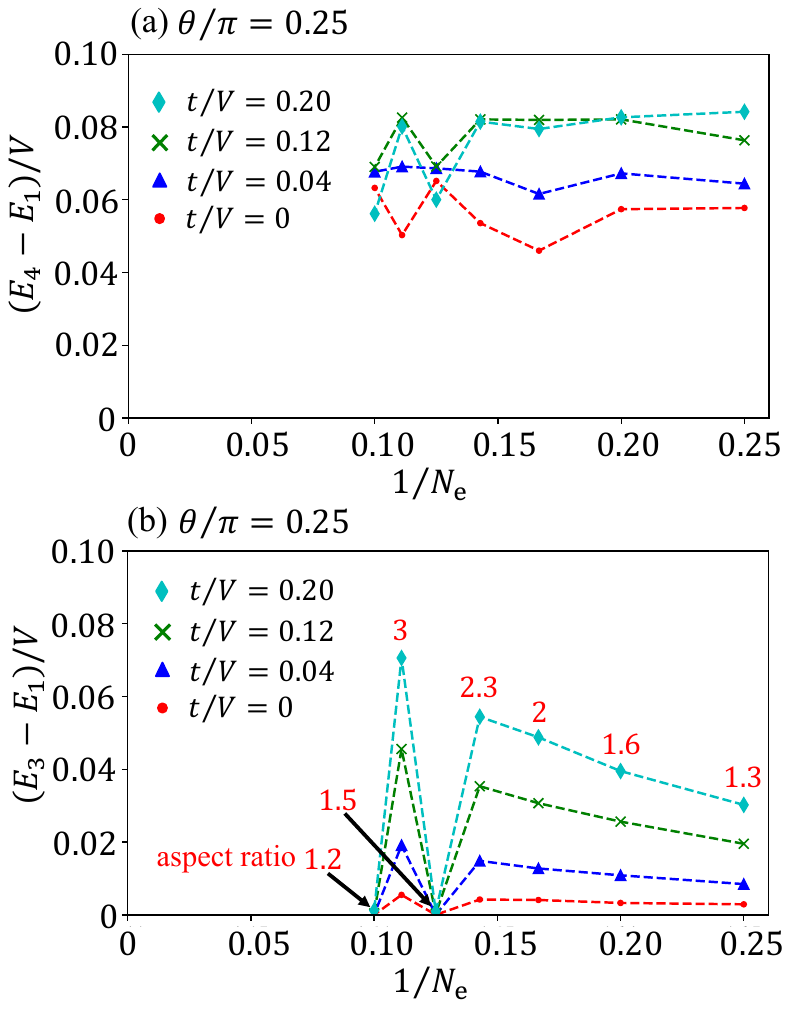}
   \caption{
    Finite size scaling of (a) $E_4 - E_1$ and (b) $E_3 - E_1$ at fixed $\theta/\pi =0.25$ for several $t/V$.
    For a given $N_{e}$, we choose the system size such that the aspect ratio is closest to unity.
   }
   \label{fss}
  \end{figure}
  \subsection{Finite size scaling and spectral flow}
  We next investigate the energy gap of the three-fold quasidegenerate ground
  state via finite-size scaling analysis. 
  In Figs.~\ref{fss}(a) and (b), we plot the energy differences
  $E_4-E_1$ and $E_3 - E_1$, respectively, as functions of the inverse of
  the electron number $1/N_e$. We present the results for several values of $t/V$. 
  (The energy is scaled by $V$, motivated by the expectation that the energy 
  gap of FCIs scales with the interaction strength.) Figure~\ref{fss}(a) 
  suggests that $E_4-E_1$ remains finite in the thermodynamic limit for all
  values of $t/V$ considered here, and is consistent with the emergence of the FCI states.

  In Fig.~\ref{fss}(b), however, the behavior of $E_3-E_1$ is more subtle
  in the thermodynamic limit. While $E_3-E_1$ appears to vanish at $t/V=0$ 
  (i.e., in the flat-band limit), the data with finite $t/V$ suggest a finite value: $E_3-E_1$ tends
  to increase as $1/N_e$ is decreased, except for the two data points. To 
  clarify this behavior, we indicate the aspect ratio $N_1/N_2$ in the figure.
  One can see that $E_3-E_1$ increases as the aspect ratio deviates from 
  unity, whereas it becomes very small when the aspect ratio is closer to one. 
  This suggests that $E_3-E_1$ may vanish in the thermodynamic limit, 
  provided the aspect ratio is kept close to unity.
  
  \begin{figure}[t]
   \centering
   \includegraphics[width=\columnwidth]{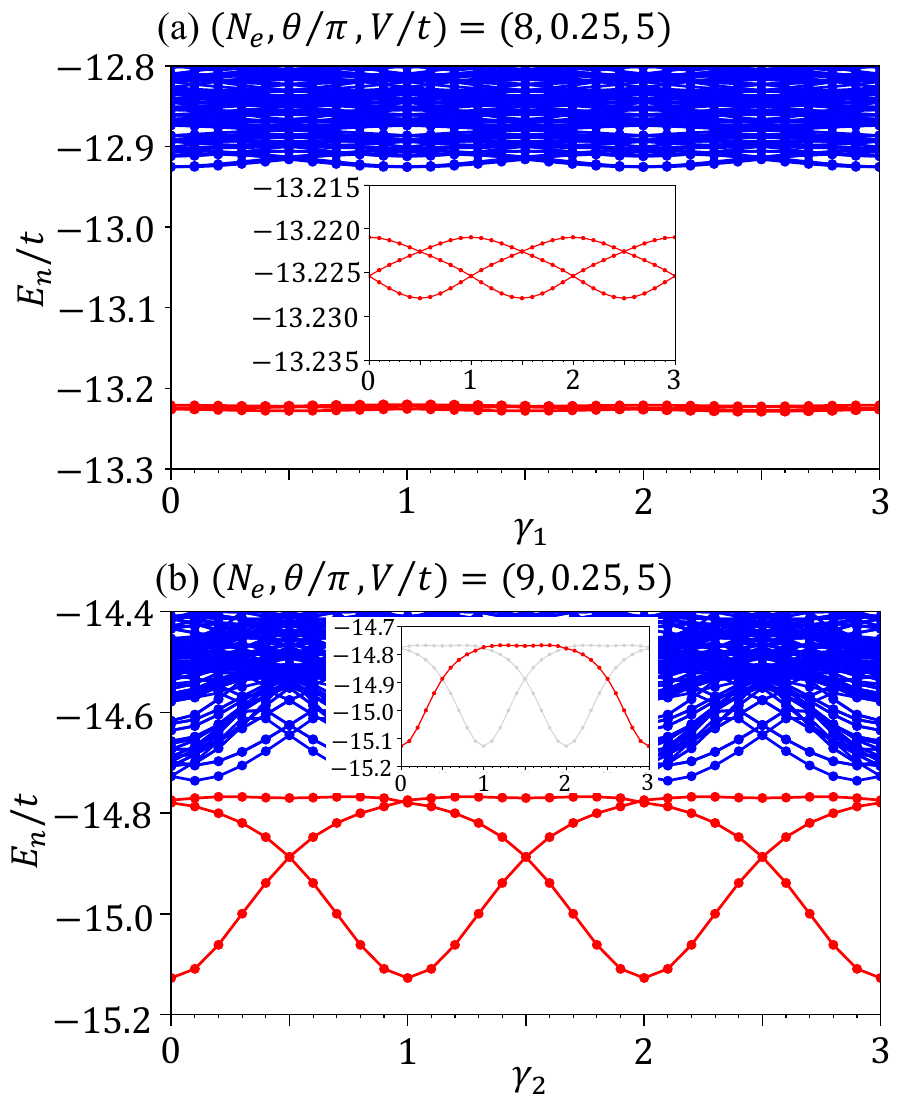}
   \caption{
   (a) Low-energy spectral flow as a function of $\gamma_1$ for
   $\theta/\pi = 0.25$, $V/t = 5$, and $N_1\times N_2 = 6\times4$.
   Here, we fix $\gamma_2 = 0$.
   Red markers indicate the lowest energies in the $(j_1,j_2) = (0,0), (2,0), (4,0)$ momentum sectors [Eq.~\eqref{fci_momentum}]. 
   The three-fold quasidegenerate ground states do not mix with higher excited states as $\gamma_1$ increases.
   After three fluxes are inserted, this configuration returns to the initial state~(inset).
   (b) The same as (a), but as a function of $\gamma_2$ for 
   $N_1\times N_2 = 9\times3$ with $\gamma_1 = 0$.
   Red markers indicate the three lowest energies in the 
   $(j_1,j_2) = (0,0)$ sector [Eq.~\eqref{fci_momentum}]. Although the three-flux periodicity is less 
   apparent, it can be seen as shown in the inset.
   }
   \label{spectral_flow}
  \end{figure}
  To further support the emergence of the FCI state, we calculate the spectral 
  flow under twisted boundary conditions~\cite{Wu2012,Regnault2011,Neupert2011,Niu1985,Kudo20192,Higashino2025}.
  This shifts the momentum index as 
  $j_{i}\rightarrow j_{i} + \gamma_{i}$ with 
  $\gamma_{i}$ real. Here, $\gamma_{i}$ can be 
  interpreted  as the number of flux quanta inserted in the $\bm{a}_i$ direction. 
  In Fig.~\ref{spectral_flow}(a), we plot the many-body energies as a
  function of $\gamma_1$ with fixed $\gamma_2=0$. The three lowest states 
  evolve into one another through level crossings while remaining separated 
  from higher excited states. Although the Hamiltonian is periodic in $\gamma_1$
  with period one, following a given branch continuously reveals an effective
  period tripling. Such behavior is consistent with the fractionalization
  expected in a $\nu=1/3$ FCI state~\cite{Neupert2011}.
  In Fig.~\ref{spectral_flow}(b), we examine a system where $E_3-E_1$ is
  comparatively large [see Fig.~\ref{fss}(b)]. The three lowest states again 
  evolve without crossing higher excited states, even though the 
  excitation gap is smaller. While the period-tripling behavior is less apparent, 
  it can become visible as illustrated in the inset.
  
  \section{Summary}
    In this paper, we investigate the emergence of the  $\nu = 1/3$ FCI 
    state in a kagome lattice with scalar spin chirality. We include the effect
    of the band dispersion to examine its impact on the stability of the FCI 
    state. We demonstrated that the FCI
    state emerges over a wide region of parameter space by calculating the 
    overlap with a representative FCI state. The emergent FCI state is characterized by
    the nearly three-fold ground-state degeneracy, the finite energy gap in the 
    thermodynamic limit, and the spectral flow with three-flux quanta 
    periodicity. Our results suggest that kagome magnets with anisotropic 
    magnetic ordering can be promising platforms for realizing FCIs.\\

  \textbf{Acknowledgements}
  This work was supported by JSPS KAKENHI Grant Nos. JP23K19036, JP25K17318, JP25H01250, JP25H00613.
  The computation in this work has been done using the facilities of the Supercomputer Center, the Institute for Solid State Physics, 
  the University of Tokyo.

  \vspace{35pt}
  \begin{figure}[h]
    \centering
    \includegraphics[width=\columnwidth]{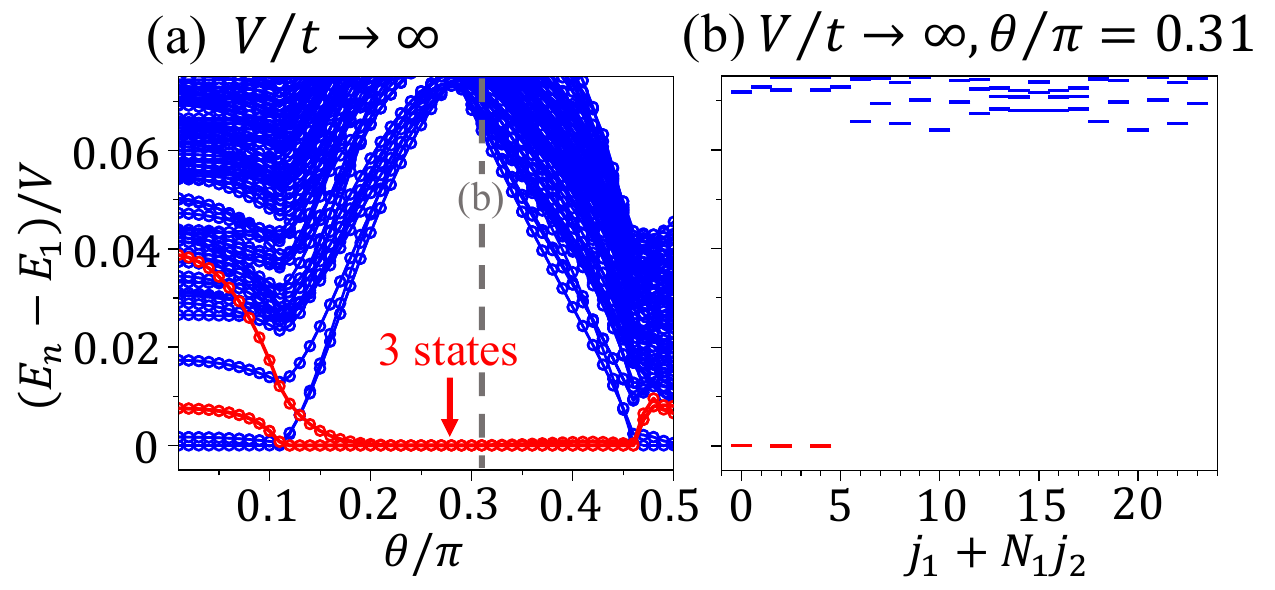}
    \caption{
      (a) Low-energy spectrum as a function of $\theta$ in the flat-band limit.
      Red markers indicate the lowest energies in the $(j_1, j_2) = (0,0), (2,0), (4,0)$ momentum sectors [Eq.~\eqref{fci_momentum}].
      The emergence of the three-fold quasidegenerate ground state is observed 
      for $0.15 \lesssim \theta/\pi \lesssim 0.45$. We plot the six lowest energies in each momentum sector.
      (b) Low-energy spectrum as a function of the total momentum index 
      $j_1 + N_1j_2$ at $\theta/\pi = 0.31$.
      The system size is $N_1\times N_2 = 6 \times 4$.
    }
    \label{flat}
  \end{figure}
  
  \section*{\centering Appendix : Emergence of the FCI state in the flat-band limit}
    \label{appx:flat-band-limit}
    Here, we show the emergence of the FCI state in the flat-band limit.
    Figure~\ref{flat}(a) plots $E_n - E_1$ as a function of $ \theta$.
    Three nearly degenerate ground states appear in the momentum sectors given by Eq.~\eqref{fci_momentum} and, 
    for the range of $0.15 \lesssim \theta/\pi \lesssim 0.45$.
    The regions outside this range are indicated by hatched gray in Fig.~\ref{overlap}.
    We also show in Fig.~\ref{flat}(b) the energy spectrum at 
    $\theta/\pi = 0.31$, studied in Ref.~\cite{Wu2012}, to verify our 
    numerical implementation.

  \newpage
  
  \nocite{*}
  \bibliographystyle{jpsj}
  \bibliography{ref}

\end{document}